\documentstyle[graphicx,pre,multicol,aps]{revtex}

\begin{document}
\title{Spatial optical solitons in nonlinear photonic crystals}

\author{Andrey A. Sukhorukov and Yuri S. Kivshar} 

\address{Nonlinear Physics Group, Research School of Physical Sciences and Engineering, Australian National University, \\ 
Canberra ACT 0200, Australia}
\maketitle

\begin{abstract}
We study spatial optical solitons in a one-dimensional {\em nonlinear photonic crystal} created by an array of thin-film nonlinear waveguides, the so-called Dirac-comb nonlinear lattice.  We analyze modulational instability of the extended Bloch-wave modes and also investigate the existence and stability of bright, dark, and ``twisted''" spatially localized modes in such periodic structures. Additionally, we discuss both similarities and differences of our general results with the simplified models of nonlinear periodic media described by the discrete nonlinear Schr\"odinger equation, derived in the tight-binding approximation, and the coupled-mode theory, valid for shallow periodic modulations of the optical refractive index. 
\end{abstract}

\pacs{PACS numbers:
42.70.Qs,
42.65.Wi,
42.65.Tg,
}

\begin{multicols}{2}
\narrowtext
\section{Introduction}

{\em Discrete spatial optical solitons} have been introduced and studied theoretically 
as spatially localized modes of periodic optical structures (see, e.g.,  Refs.~\cite{theory,theory1a,theory1b,theory1c} and also a review paper~\cite{lederer}), and they have recently been observed experimentally in arrays of nonlinear single-mode optical waveguides~\cite{exp}.  A standard theoretical approach in the study of the discrete spatial optical solitons is based on the derivation of an effective discrete nonlinear Schr\"odinger (DNLS) equation~\cite{theory}, and the analysis of its stationary localized solutions - discrete localized modes~\cite{lederer}.  In the solid-state physics, the similar approach is known as {\em the tight-binding approximation} which, in application to optical waveguide arrays, corresponds to the case of {\em weakly coupled} fundamental modes excited in each waveguide of the array. The analogous concepts appear in other fields such as the nonlinear dynamics of the Bose-Einstein condensates in optical lattices~\cite{smerzi}.

On the other hand, weak nonlinear effects in optical fibers with a periodic modulation of the refractive index (often called {\em optical grating}) are well studied in the framework of another approach, {\em the coupled-mode theory}. The coupled-mode theory is based on a decomposition of the electric field into the forward and backward propagating components, under the condition of the Bragg resonance. Such an approach is usually applied to analyze nonlinear localized waves in the systems with a weakly modulated optical refractive index known as {\em gap (or Bragg) solitons}~\cite{gap_review}, and such gap solitons are known to appear in other fields~\cite{zobay}. 

Thus, the theory of spatial and temporal optical solitons in periodic structures developed so far is based on one of the two approaches, the DNLS equation or the coupled-mode theory. However, real experiments in the nonlinear guided-wave optics are conducted in the periodic structures of more complicated geometries and under the conditions when none of those approximations are valid. In such a case the applicability of the tight-binding approach and the corresponding discrete equations, from one hand, and the coupled-mode theory, from the other hand, become questionable, especially for the analysis of the linear stability of nonlinear localized modes. Therefore, a consistent theory of nonlinear effects and localized modes in periodic media is still missing.

One of the main features of wave propagation in periodic structures (which follows from the Floquet-Bloch theory) is the existence of a set of forbidden band gaps in the transmission spectrum. Therefore, the nonlinearly-induced wave localization can become possible in each of these gaps. However, the effective DNLS equation derived in the tight-binding approximation describes only one transmission band surrounded by two semi-infinite band gaps and, therefore, a fine structure of the band-gap spectrum associated with the wave transmission in a periodic medium is lost. On the other hand, the coupled-mode theory of gap solitons~\cite{gap_review} describes only the modes localized in an isolated narrow gap, and it does not allow to consider simultaneously the gap modes and conventional guided waves localized due to the total internal reflection. The complete band-gap structure of the transmission spectrum and simultaneous existence of localized modes of different types are very important issues in the analysis of stability of nonlinear localized modes~\cite{our_PRL}. Such an analysis is especially important for the theory of nonlinear localized modes and nonlinear waveguides in realistic models of nonlinear photonic crystals (see, e.g., the recent paper~\cite{PBG} and references therein).

In this paper, we consider a simple model of nonlinear periodic layered media where a periodic optical structure is formed by an array of thin-film nonlinear waveguides embedded into an otherwise linear dielectric medium (see also Ref.~\cite{gera}). Such a structure can be regarded as a nonlinear analog of the so-called Dirac comb lattice~\cite{comb}, where the effects of the linear periodicity and band-gap spectrum are taken into account explicitly, whereas nonlinear effects enter the corresponding matching conditions allowing a direct analytical study.

We analyze nonlinear localized modes in an infinite structure consisting of a periodic array of nonlinear waveguides, similar to the geometry of the experiments with discrete optical solitons~\cite{exp}. First, we study modulational instability of extended modes in both self-focusing and self-defocusing regimes. Then, we discuss different types of nonlinear localized modes (such as bright, dark and ``twisted'' spatial solitons) and also analyze numerically their linear stability.  We emphasize both similarities and differences between our results and the results obtained in the framework of the DNLS equation and the continuous coupled-mode theory.

\section{General Approach} \label{sect:general}
\subsection{Model} \label{sect:model}

\begin{figure}
\centerline{\includegraphics[width=7.5cm,clip]{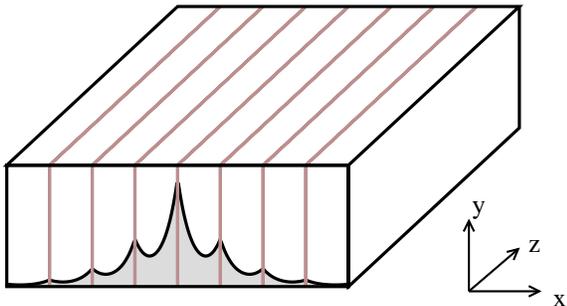}}
\caption{ \label{fig:structure}
Array of thin-film nonlinear waveguides embedded into a linear slab waveguide. Gray shading shows a profile of a nonlinear localized mode.
}
\end{figure}

We consider the electromagnetic waves propagating along the $Z$-direction of a slab-waveguide structure created by a periodic array of thin-film nonlinear waveguides (see Fig.~\ref{fig:structure}). Assuming that the field structure in the $Y$ direction is defined by the linear guided mode of the slab waveguide ${\cal E}(Y; X)$, we separate the dimensions presenting the electric field as $E(X,Z)\,{\cal E}(Y; X)$. Then, the evolution of the complex field envelope $E(X,Z)$ is governed by the nonlinear Schr\"odinger (NLS) equation,
\begin{equation} \label{eq:nls_dim} 
       i \frac{\partial E}{\partial Z} 
       + D \frac{\partial^2 E}{\partial X^2} 
       + \varepsilon(X) E + g(X) |E|^2 E = 0,
\end{equation}
where $D$ is the diffraction coefficient ($D>0$). The phase velocity of the guided waves is defined by the function $\varepsilon(X)$, whereas $g(X)$ characterizes the Kerr-type nonlinear response of the layers. We assume that either the function $\varepsilon(X)$ or $g(X)$ (or both of them) is periodic in $X$, i.e. it describes the periodic layered structure similar to the so-called transverse Bragg waveguides created by the nonlinear thin-film multilayer structures~\cite{grebel,nabiev} or the impurity band in a deep photonic band gap~\cite{lan}.  

In order to reduce the number of physical parameters, we normalize Eqs.~(\ref{eq:nls_dim}) as follows: 
$E(X,Z) = \psi(x,z) E_0 e^{i \; \overline{\varepsilon} Z}$, where $\overline{\varepsilon}$ is the mean value of the function $\varepsilon(X)$, $x=X/d$ and $z=Z D / d^2$ are the dimensionless coordinates, $d$ and $E_0$ are the characteristic transverse scale and field amplitude, respectively. Then, the normalized nonlinear equation has the form
\begin{equation} \label{eq:nls}  
     i \frac{\partial \psi}{\partial z} 
     + \frac{\partial^2 \psi}{\partial x^2}  
     + {\cal F}(I; x) \psi = 0,
\end{equation}
where the real function ${\cal F}(I; x) = d^2 D^{-1} [ \varepsilon(X) - \overline{\varepsilon} + g(X) I |E_0|^2]$ describes both {\em nonlinear} and {\em periodic} properties of the layered medium, and $I \equiv |\psi|^2$ is the normalized local wave intensity. We note that the system~(\ref{eq:nls}) is Hamiltonian, and for spatially localized solutions it conserves the total power, 
\[
   P = \int_{-\infty}^{+\infty} |\psi(x,z)|^2 \; dx .
\]    

At this point, it is important to mention that Eq.~(\ref{eq:nls}) describes the beam evolution in the framework of the so-called {\em parabolic approximation}, valid for the waves propagating mainly along the $z$ direction (see also Ref.~\cite{our_old_pre} and discussions therein).  In other words, the characteristic length of the beam distortion due to both diffraction and refraction along the $z$ axis should be much larger than the beam width in the transverse direction $x$. This leads to the condition of {\em a weakly modulated periodicity}, $|\varepsilon(X) - \overline{\varepsilon}| \ll |\overline{\varepsilon}|$. 

We look for stationary localized solutions of the normalized equation (\ref{eq:nls}) in the standard form
\begin{equation} \label{eq:lmode}
  \psi(x, z) = u(x; \beta) e^{i \beta z}, 
\end{equation}
where $\beta$ is the propagation constant, and the amplitude function $u(x; \beta)$ satisfies the stationary nonlinear equation:
\begin{equation} \label{eq:u0_inh}
  - \beta u + \frac{d^2 u}{d x^2} + {\cal F}(I; x) u = 0.
\end{equation}
If there is no energy flow along the transverse direction $x$, then the function $u(x)$ is real, up to a constant phase which can be removed by a coordinate shift $z \rightarrow z-z_0$. This is always the case for spatially localized solutions with vanishing asymptotics, $u(x\rightarrow\pm\infty)=0$.

To simplify our analysis further, we assume that the linear periodicity is associated only with the presence of an array of the thin-film waveguides, and define the response function in the model Eq.~(\ref{eq:nls}) as follows
\begin{equation} \label{eq:nls_array_resp}
   {\cal F}(I; x) = \sum_{n=-\infty}^{+\infty} 
                     (\alpha + \gamma I) \delta(x - h n),
\end{equation}
where $h$ is the spacing between the neighboring thin-film waveguides (the lattice period), and $n$ is integer. The total response of the thin-film layers is approximated by the delta-functions, and the real parameters $\alpha$ and $\gamma$ describe both {\em linear} and {\em nonlinear} properties of the layer, respectively. Without loss of generality, the nonlinear coefficient~$\gamma$ can be normalized to unity, so that $\gamma=+1$ corresponds to {\em self-focusing} and $\gamma=-1$ to {\em self-defocusing} nonlinearity. The linear coefficient ($\alpha>0$) defines the low-intensity response, and it characterizes the corresponding coupling strength between the waveguides. Model~(\ref{eq:nls}),(\ref{eq:nls_array_resp}) can be regarded as a nonlinear analog of the Dirac comb lattice, earlier studied in application to the photonic crystals in the linear regime only~\cite{comb}.
 
\subsection{Dispersion Properties and Discrete Equations}
            \label{sect:array_discrete}

Following the path outlined in Ref.~\cite{our_old_pre}, we present the stationary modes defined by Eqs.~(\ref{eq:u0_inh}),(\ref{eq:nls_array_resp}) as a sum of the counter-propagating waves in each of the linear slab waveguides,
\begin{equation} \label{eq:prof_ab}
   u(x) = a_n e^{- \mu (x-n h)} + b_n e^{+ \mu (x-n h)} ,
\end{equation}
where $n h \le x \le (n+1) h$. Then, we express the coefficients $a_n$ and $b_n$ in terms of the wave amplitudes at the nonlinear layers, $u_n = u(h n)$, 
\begin{equation} \label{eq:prof_ab_un}
 \begin{array}{l} 
 { \displaystyle
   a_n = \frac{u_n e^{\mu h} - u_{n+1}}{2\; {\rm sinh}\,(\mu h)}, 
 } \\*[9pt] { \displaystyle
   b_n = u_n - a_n ,
 } \end{array}
\end{equation}
where $\mu(\beta) = \sqrt{\beta}$. Finally, we substitute Eqs.~(\ref{eq:prof_ab}),(\ref{eq:prof_ab_un}) into Eqs.~(\ref{eq:u0_inh}),(\ref{eq:nls_array_resp}), and find that the normalized amplitudes $U_n = \sqrt{|\xi \gamma|} u_n$ satisfy a stationary form of the DNLS equation:
\begin{equation} \label{eq:dnls}
  \eta U_n + ( U_{n-1} + U_{n+1} ) + \chi |U_n|^2 U_n = 0 ,
\end{equation}
where $\chi = {\rm sign}(\xi \gamma)$, and 
\begin{equation} \label{eq:dnls_param}
 \begin{array}{l} 
 { \displaystyle
   \eta(\beta) = - 2\; {\rm cosh}( \mu h ) 
           + \alpha \xi(\beta) ,
 } \\*[9pt] { \displaystyle
   \xi(\beta) = {\rm sinh}( \mu h ) / \mu . 
 } \end{array}
\end{equation}

\begin{figure}[H]
\centerline{\includegraphics[width=7.5cm,clip]{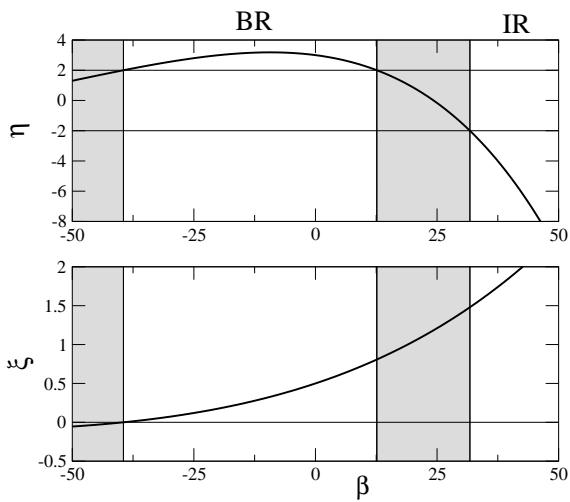}}
\vspace*{1mm}
\caption{ \label{fig:dispers}
Characteristic dependencies of the parameters $\eta$ and $\xi$ on the propagation constant $\beta$. Gray shading marks the transmission bands.
The lattice parameters are $h=0.5$ and $\alpha=10$.
}
\end{figure}

Linear solutions of Eq.~(\ref{eq:dnls}) (when the term $\sim \chi |U_n|^2$ vanishes) have the form $U_n = U_0 e^{i K n}$, where $K = \pm {\cos}^{-1} (-\eta/2)$ is the wave number. Therefore, extended linear solutions with real $K$ can exist for $|\eta| \le 2$. On the other hand, nonlinear localized modes with exponentially decaying asymptotics can appear only for $|\eta| > 2$ (when $K$ is imaginary), this condition defines the {\em band-gap structure of the spectrum}. Characteristic dependencies of $\eta$ and $\xi$ vs. the propagation constant $\beta$ are presented in Fig.~\ref{fig:dispers}, where the bands are shown by a gray shading. The first (semi-infinite) band gap corresponds to the total {\em internal reflection} (IR). On the other hand, at smaller $\beta$ the {\em spectrum band gaps} appear due to the resonant Bragg-type reflection (BR) from the periodic structure.

\subsection{Linear Stability Analysis}
            \label{sect:stability}

To study {\em linear stability} of localized modes, we should consider the evolution of small-amplitude perturbations of the localized state presenting the solution in the form
\begin{equation} \label{eq:lperturb}
 \psi (x,t) 
 = \left\{ u(x) + v(x) e^{i \Gamma z} 
                + w^{\ast}(x) e^{-i \Gamma^{\ast} z} 
   \right\} e^{i \beta z} .
\end{equation}
From Eqs.~(\ref{eq:nls}) and~(\ref{eq:nls_array_resp}), we obtain the linear eigenvalue problem for small $v(x)$ and $w(x)$:
\begin{eqnarray} \label{eq:Leigen}
  \begin{array}{l}
   { \displaystyle
      -(\beta+\Gamma) v + \frac{d^2 v}{d x^2} 
   } \\*[9pt] { \displaystyle \quad
      + \sum_{n=-\infty}^{+\infty} 
                        \left[(\alpha + 2 \gamma |u_n|^2) v 
                             + \gamma u_n^2 w  \right]
                     \delta(x - h n) = 0 , 
   } \\*[9pt] { \displaystyle
      -(\beta-\Gamma) w + \frac{d^2 w}{d x^2} 
   } \\*[9pt] { \displaystyle \quad
      + \sum_{n=-\infty}^{+\infty} 
                    \left[(\alpha + 2 \gamma |u_n|^2)  w 
                             + \gamma (u_n^{\ast})^2 v \right]
                     \delta(x - h n) = 0 .
   } \end{array}
\end{eqnarray}
After representing the fields as sums of the counter-propagating waves (see Sec.~\ref{sect:array_discrete}), we reduce Eqs.~(\ref{eq:Leigen}) to a set of the discrete equations for the amplitudes at the layers:
\begin{equation} \label{eq:Leigen_discrete}
  \begin{array}{l}
   { \displaystyle
      \eta(\beta+\Gamma) v_n 
      + ( v_{n-1} + v_{n+1} ) + 
   } \\*[9pt] { \displaystyle \qquad\qquad
      + \gamma \xi(\beta+\Gamma) \left[ 2 |u_n|^2 v_n 
                                 + u_n^2 w_n \right] = 0 ,
   } \\*[9pt] { \displaystyle
      \eta(\beta-\Gamma) w_n 
      + ( w_{n-1} + w_{n+1} ) + 
   } \\*[9pt] { \displaystyle \qquad\qquad
      + \gamma \xi(\beta-\Gamma) \left[ 2 |u_n|^2 w_n 
                                 + (u_n^{\ast})^2 v_n \right] = 0 .
   } \end{array}
\end{equation}
In general, the solutions of this eigenvalue problem fall into one of the following categories: (i)~{\em internal modes} with real eigenvalues that describe periodic oscillations (``breathing'') of the localized state, (ii)~{\em instability modes} that correspond to purely imaginary eigenvalues with ${\rm Im}\Gamma < 0$, and (iii)~{\em oscillatory unstable modes} that appear when the eigenvalues are complex (and ${\rm Im}\Gamma < 0$). Additionally, there can exist decaying modes when ${\rm Im}\Gamma > 0$. However, from the structure of Eqs.~(\ref{eq:Leigen_discrete}) it follows that the eigenvalue spectrum is invariant with respect to the transformation $\Gamma \rightarrow \pm \Gamma^{\ast}$, if all the amplitudes $u_n e^{i \phi}$ are real, where $\phi$ is an arbitrary constant phase. In such a case exponentially growing and decaying modes always coexist, and the latter do not significantly affect the wave dynamics.

It is important to note that the grating properties are expressed through the functions $\eta(\beta)$ and $\xi(\beta)$ [see Eq.~(\ref{eq:dnls_param})] in both stationary [Eq.~(\ref{eq:dnls})] and perturbation [Eq.~(\ref{eq:Leigen_discrete})] equations. Therefore, the functions $\eta(\beta)$ and $\xi(\beta)$ fully characterize the existence and linear stability properties of localized and extended solutions of the original model~(\ref{eq:nls}),(\ref{eq:nls_array_resp}) with stationary intensity profiles.

\section{Approximate Models}  \label{sect:approximate}
\subsection{Tight-Binding Approximation}  \label{sect:tight}

When each of the thin-film waveguides of the periodic structure supports a fundamental mode that weakly overlaps with the similar mode of the neighboring waveguide, the modes become weakly coupled via a small change of the refractive index in the waveguides. Then, the mode properties can be analyzed in the framework of the so-called {\em tight-binding approximation}, well developed for the problems of solid-state physics~\cite{tight}.

In order to employ this approximation in our case, first we analyze the properties of a single thin-film waveguide in the linear regime and solve Eq.~(\ref{eq:u0_inh}) with ${\cal F}(I; x) = \alpha \delta(x)$ to find the spatial profile of a linear guided mode,
\[
   u_s(x) = \exp\left({- \alpha |x| / 2}\right) . 
\]
and the corresponding value of its propagation constant, $\beta_s = \alpha^2 / 4$.  Second, we consider the interaction between the waveguides in the array
assuming that the total field can be presented as a superposition of slightly perturbed waves localized at the isolated waveguides. Specifically, we assume that the propagation constant remains close to its unperturbed value $\beta_s$ (i.e. $|i d \psi / d t + \beta_s \psi| \ll |\beta_s \psi|$), and neglect small variations in the spatial profiles of the localized modes. Then, we seek a general solution of Eqs.~(\ref{eq:nls}),(\ref{eq:nls_array_resp}) in the form 
\begin{equation} \label{eq:tight_field}
  \psi(x, z) = \sum_{n=-\infty}^{+\infty} \psi_n(z) u_s(x - n h) .
\end{equation}
In such an approximation, the wave evolution is characterized by the amplitude functions $\psi_n(z)$ only. In order to find the corresponding evolutionary equations, we substitute Eq.~(\ref{eq:tight_field}) into Eqs.~(\ref{eq:nls}) and~(\ref{eq:nls_array_resp}) and, multiplying the resulting equation by $u_s(x-m h)$, integrate it over the transverse profile. According to the original assumption of weakly interacting waveguides (valid for $\alpha h \gg 1$ and $|\gamma| |\psi_n(z)|^2 \ll \alpha$), in the lowest-order approximation we put $|\psi(n h,z)|^2  \simeq |\psi_n(z)|^2$ and neglect the overlap integrals, $\int_{-\infty}^{+\infty} u_s(x-n h) u_s(x-m h) dx$, with $|n-m|>1$, which produce the terms of higher orders.  Finally, we derive 
a system of coupled discrete equations for the field amplitudes at the nonlinear layers (up to small perturbations, since $\psi(n h, z) = \psi_n(z) + O(e^{-\alpha h/2})$):
\begin{equation} \label{eq:dnls_tight}
  \begin{array}{l}
   { \displaystyle
      i \frac{d \psi_n}{d z} + \beta_s \psi_n 
      + \frac{\alpha^2}{2} e^{-\alpha h/2} ( \psi_{n-1} + \psi_{n+1} ) 
   } \\*[9pt] { \displaystyle \qquad \qquad \qquad \qquad
      + \frac{\gamma \alpha}{2} |\psi_n|^2 \psi_n = 0 .
   } \end{array}
\end{equation}
Stationary solutions of Eq.~(\ref{eq:dnls_tight}) have the form $\psi_n = u_n e^{i \beta z}$, where the amplitudes $u_n$ are given by Eq.~(\ref{eq:dnls}) with 
\begin{equation} \label{eq:dnls_tight_dispers}
   \eta = -\frac{2}{\alpha^2} \left(\beta-\frac{\alpha^2}{4}\right) e^{-\alpha h / 2},  
          \quad
   \xi = \frac{1}{\alpha} e^{-\alpha h / 2} .
\end{equation}
Relations~(\ref{eq:dnls_tight_dispers}) can be found as a series expansion of the original dispersion relations~(\ref{eq:dnls_param}) near the edges of the first transmission band, in the limit $\alpha h \gg 1$. 

\subsection{Coupled-Mode Theory}    \label{sect:coupled}

Now we consider the opposite limit $\alpha h \ll 1$, when the first Bragg-reflection gap is narrow, i.e. $(\beta_1 - \beta_2) \ll |\beta_{1,2}|$, where $\beta_{1,2}$ are the propagation constants at the gap edges, defined by the condition $\eta(\beta_{1,2}) \equiv 2$. From Eq.~(\ref{eq:dnls_param}) it follows that $\beta_2 = - (\pi/h)^2$ and $\beta_1 \simeq \beta_2 + 2 \alpha / h + O(\alpha^{3/2})$.
To find solutions close to the BR gap, we present the total field in the form
\begin{equation} \label{eq:coupled_bloch}
   \psi(x,z) = a_1(x,z) u_b(x; \beta_1)  
               + a_2(x,z) u_b(x; \beta_2) ,
\end{equation}
where $a_j$ are unknown nonlinear amplitudes, and $u_b(x; \beta_{1,2})$ are the linear Bloch functions which satisfy Eqs.~(\ref{eq:u0_inh}) and~(\ref{eq:nls_array_resp}) at $\gamma = 0$. The Bloch functions can be found in an explicit form: $u_b(x; \beta_2) = \sin(x \pi / h)$, and $u_b(x+n h; \beta_1) = (-1)^n \sin[ (h/2-x) \sqrt{|\beta_1|}]$ for $0 \le x \le h$. Note that the field amplitudes at the layers are $u_n \sim (-1)^n a_1(n h)$, since $u_b(n h; \beta_2) \equiv 0$.

In order to find the equations for the amplitudes $a_j(x,z)$, we substitute Eq.~(\ref{eq:coupled_bloch}) into the original model Eqs.~(\ref{eq:nls}) and~(\ref{eq:nls_array_resp}). Next, we use the fact that the gap is narrow, and close to its edges the Bloch functions are weakly modulated, i.e. 
$|\partial a_{1,2} / \partial x| \ll |a_{1,2} / h|$. This assumption allows us to keep only the lowest-order terms. Then, we multiply the resulting equation by $u_b(x; \beta_{1,2})$ and integrate it over one grating period. Finally, the coupled equations for the modulation amplitudes are 
\begin{equation} \label{eq:coupled}
  \begin{array}{l}
   { \displaystyle
      i \frac{\partial a_1}{\partial z} 
      + \beta_1 a_1 
      + \frac{2 \pi}{h} \frac{\partial a_2}{\partial x} 
      + \gamma \frac{2}{h} |a_1|^2 a_1 
      = 0 ,
   } \\*[9pt] { \displaystyle
      i \frac{\partial a_2}{\partial z} 
      + \beta_2 a_2 
      - \frac{2 \pi}{h} \frac{\partial a_1}{\partial x} 
      = 0 ,
   } \end{array}
\end{equation}

Equations (\ref{eq:coupled}) derived above allow a direct comparison between the coupled-mode theory and the general results for the stationary localized solutions for which $a_{1,2}(x,z) = a_{1,2}(x) e^{i \beta z}$. Additionally, since the functions $a_j$ are weakly modulated, the spatial derivatives can be approximated by finite differences between the amplitudes at the layers (note that such an approximation is only valid for  $\beta \simeq \beta_{1,2}$). After simple algebra  we obtain a discrete equation for the field amplitudes at the layers which has the form of the DNLS equation~(\ref{eq:dnls}) with 
\begin{equation} \label{eq:coupled_dispers}
  \begin{array}{l}
   { \displaystyle
      \eta (\beta) = 2 - \frac{h^4}{4 \pi^2} (\beta-\beta_1) (\beta-\beta_2),
   } \\*[9pt] { \displaystyle
      \xi (\beta) = \frac{h^3}{2 \pi^2} (\beta-\beta_2) .
   } \end{array}
\end{equation}
Similar to the case of the tight-binding approximation, the dispersion relations~(\ref{eq:coupled_dispers}) can be found from the general result~(\ref{eq:dnls_param}) by performing a series expansion near the band edge value $\beta_2$. 

\subsection{Two-Component Discrete Model}
            \label{sect:superlattice}

The principal limitations of both the tight-binding approximation and the coupled-mode theory is explained by the fact that they are both valid in local narrow regions of the general band-gap structure and under special assumptions. Indeed, these two approaches are applicable when the dimensionless parameter $\alpha h$ is either small or large, and none of those approaches covers the intermediate cases. While a general study should rely on the numerical solutions with the exact dispersion relations~(\ref{eq:dnls_param}), it is useful to consider a simplified model which can (at least, qualitatively) describe the wave properties close to the BR gap ($\beta \simeq \beta_1$) in the transitional region, being valid for $\alpha h \simeq 1$ as well. To achieve this goal, {\em we extend the tight-binding approximation and the corresponding DNLS} equation introduced above in Sec.~\ref{sect:tight}. 

We note that Eq.~(\ref{eq:dnls_tight}) can be considered as a rough discretization of the original model~(\ref{eq:nls}), with only one node per grating period located at $x=n h$. Then, the natural generalization is to include additional nodes located between the nonlinear layers at the positions $x=(n+1/2) h$. Since the refractive indices at the node position are now different, we obtain {\em a new system of coupled discrete equations} which correspond to a two-component superlattice:
\begin{equation} \label{eq:2atom}
  \begin{array}{l}
   {\displaystyle
      i \frac{d \psi_n}{d z} + \beta_1 \psi_n 
      + \rho_1 ( \psi_{n-1/2} + \psi_{n+1/2}) 
   } \\*[9pt] {\displaystyle \qquad \qquad \qquad \qquad \qquad \qquad
      + \widetilde{\gamma} |\psi_n|^2 \psi_n = 0,
   } \\*[9pt] {\displaystyle
      i \frac{d \psi_{n+1/2}}{d z} + \beta_2 \psi_{n+1/2} 
      + \rho_2 (\psi_{n} + \psi_{n+1}) = 0,
   } \end{array}
\end{equation}
We find that, similar to the general case, the stationary mode profiles can be expressed in terms of the amplitudes $u_{n}$, which satisfy the normalized DNLS Eq.~(\ref{eq:dnls}) with the following parameters:
\begin{equation} \label{eq:2atom_dispers}
  \begin{array}{l}
   {\displaystyle
      \eta (\beta) = 2 - (\rho_1 \rho_2)^{-1} (\beta-\beta_1) (\beta-\beta_2),
   } \\*[9pt] {\displaystyle
      \xi (\beta) = \widetilde{\gamma} \gamma^{-1} (\rho_1 \rho_2)^{-1} 
           (\beta-\beta_2).
   } \end{array}
\end{equation}

One can immediately see that our new model~(\ref{eq:2atom}) describes an effective system with a semi-infinite IR gap and a BR gap of a finite width. We note that, although dispersion relations in Eqs.~(\ref{eq:2atom_dispers}) and~(\ref{eq:coupled_dispers}) look similar, the latter relation is only valid for $\beta \simeq \beta_{1,2}$, so that the coupled-mode theory describes only a single isolated Bragg-reflection gap. Therefore, the model~(\ref{eq:2atom}) provides an important generalization of the DNLS theory, which also has a wider applicability than the coupled-mode theory.

Because the model~(\ref{eq:2atom}) describes an extra gap in the transmission spectrum, a comparison with the original model becomes more complicated. We choose the model parameters $\rho_j$ and $\widetilde{\gamma}$ in order to match the dispersion relations in the vicinity of $\beta_1$, and satisfy a relation between $u_n$ and $u_{n+1/2}$ following from Eqs.~(\ref{eq:prof_ab}),(\ref{eq:prof_ab_un}):
\begin{equation} \label{eq:2atom-param}
  \begin{array}{l}
   { \displaystyle
      \rho_1 = -2 \cosh(\sqrt{\beta_1} h /2) / (d \eta / d \beta)_{\beta_1}
   } \\*[9pt] { \displaystyle
      \rho_2 = (\beta_1 - \beta_2) / [ 2 \cosh(\sqrt{\beta_1} h /2)],
   } \\*[9pt] { \displaystyle
      \widetilde{\gamma} = - \gamma \xi(\beta_1) / 
                              (d \eta / d \beta)_{\beta_1} ,
   } \end{array}
\end{equation}
where the functions $\eta(\beta)$ and $\xi (\beta)$ should be calculated according to Eq.~(\ref{eq:dnls_param}).

The three different approximate models discussed above allow a simple analysis of the limiting cases, and they will be used below in calculating different properties of the extended and localized modes.

\section{Modulational Instability}
            \label{sect:array_mi}
\subsection{General Analysis} \label{sect:mi_general}

First, we analyze the properties of the simplest extended (plane-wave) solutions of the model~(\ref{eq:nls}) and~(\ref{eq:nls_array_resp}) which have equal intensities at the nonlinear layers, $I_0 = |u_n|^2 = {\rm const}$, and correspond to {\em the first transmission band}. These solutions have the form of the so-called Bloch waves (BWs) $u_n = u_0 e^{i K n}$, where the wave number $K$ is selected in the first Brillouin zone, $|K| \le \pi$. Using Eq.~(\ref{eq:dnls}), we find the dispersion relation as $K = \pm \cos^{-1}[\eta(\beta; \tilde{\alpha})]$, where $\tilde{\alpha} = (\alpha + \gamma I_0^2)$ defines the layers response modified by nonlinearity. Since the transmission bands are defined by the condition $|\eta| < 2$ (see Sec.~\ref{sect:array_discrete}), the band structure shifts as intensity increases. Indeed, by resolving the dispersion relation we determine a relation between the propagation constant and the wave intensity, 
\begin{equation} \label{eq:BW_I0}
   I_0(\beta) = -\frac{[2 \cos \, K + \eta(\beta)]}{\gamma \xi(\beta)}. 
\end{equation}
Since in the first transmission band $\beta>-(\pi/h)^2$, we have
$\xi(\beta)>0$ (see also Fig.~\ref{fig:dispers}), and it follows from Eq.~(\ref{eq:dnls_param}) that the propagation constant $\beta$ increases at higher intensities in a self-focusing medium ($\gamma>0$), and decreases in a self-defocusing medium ($\gamma<0$).

One of the main problems associated with the nonlinear BW modes is their {\em instability to periodic modulations} of a certain wavelength, known as {\em modulational instability} (see also~\cite{mi_bec}). In order to describe the stability properties of the periodic BW solutions, we analyse the evolution of weak perturbations described by the eigenvalue problem~(\ref{eq:Leigen_discrete}). Due to periodicity of the background solution $u(x)$, it follows from the Bloch theorem that the eigenmodes of Eq.~(\ref{eq:Leigen_discrete}) should also be periodic, i.e. $v(x+h)=v(x) e^{i (q+K)}$ and $w(x+h) =  w(x) e^{i (q-K)}$. Then, we obtain the following solvability condition: 
\begin{equation} \label{eq:mi_cond}
 \begin{array}{l}
  {\displaystyle
    \left[ \eta(\beta+\Gamma) 
           + 2 \gamma \xi(\beta+\Gamma) I_0 
           + 2 \cos(q + K) \right] 
  } \\*[9pt] {\displaystyle
    {\rm x} \left[ \eta(\beta-\Gamma) 
           + 2 \gamma \xi(\beta-\Gamma) I_0 
           + 2 \cos(q - K) \right]
  } \\*[9pt] {\displaystyle
    = \gamma^2 \xi(\beta+\Gamma) \xi(\beta-\Gamma) I_0^2.
  } \end{array}
\end{equation}
Possible eigenvalues $\Gamma$ are determined from the condition that the spatial modulation frequencies $q$, which are found from Eq.~(\ref{eq:mi_cond}), are all real.
Therefore, the eigenvalue spectrum consists of bands, and the instability growth rate can only change continuously from zero to some maximum value. 
We also note that the spectrum possesses a symmetry $\Gamma \rightarrow \pm \Gamma^{\ast}$, and it is sufficient to study only the solutions with ${\rm Re}\,\Gamma \ge 0$.

\begin{figure}[H]
\centerline{\includegraphics[width=8cm,clip]{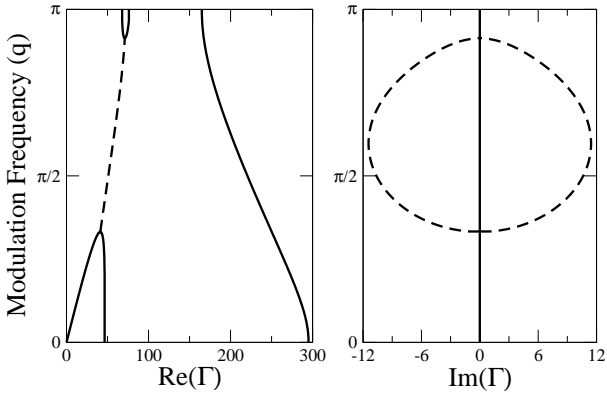}}
\vspace*{1mm}
\caption{ \label{fig:mi_freq_bifurc}
Modulation frequency $q$ vs. real (left) and imaginary (right) parts of the perturbation eigenvalue $\Gamma$, for the staggered BW modes in a self-focusing medium ($K = \pi$, $\beta = 7$, $\alpha=3$, $h=0.5$, $\gamma=+1$). Solid lines~--- stable (${\rm Im}\, \Gamma=0$) and dashed~--- unstable (${\rm Im}\, \Gamma \ne 0$).
}
\end{figure}

\subsection{Stability of Staggered and Unstaggered Modes}

In what follows, we consider two characteristic cases of the stationary nonlinear BW modes, when they are (i)~unstaggered ($K=0$) or (ii)~staggered ($K=\pi$). To describe a transition from purely real to complex linear eigenvalues, we consider the function 
\[
  Q(\Gamma) = \cos[q(\Gamma)] ,
\]
defined from Eq.~(\ref{eq:mi_cond}), 
and extend the solution from the real axis to the complex plane by writing a series expansion:
$Q({\rm Re} \Gamma + i {\rm Im} \Gamma) 
= Q({\rm Re} \Gamma) 
+ Q^{\prime}({\rm Re} \Gamma) (i {\rm Im} \Gamma) 
+ (1/2) Q^{\prime\prime}({\rm Re} \Gamma) (i {\rm Im} \Gamma)^2
+ O( {\rm Im} \Gamma^3 )$, 
where the prime denotes differentiation with respect to the argument.
Since the modulation frequency should remain real, the second term in the series expansion should vanish. Then,
we conclude that complex eigenvalues only appear at the critical points, where
\begin{equation} \label{eq:mi_bif}
   Q^{\prime}(\Gamma) = 0. 
\end{equation}
Therefore, (in)stability can be predicted by studying the function $Q(\Gamma)$ on the real axis only, and then extending the solution to the complex plane at the critical points (if they are present) determined by Eq.~(\ref{eq:mi_bif}). The real modulation frequencies are found as $q = \cos^{-1}(Q)$ in the interval $-1 \le Q \le 1$, therefore {\em MI only appears at the critical points where $q^{\prime}(\Gamma) = 0$ or $q = 0, \pi$}, see an example in Fig.~\ref{fig:mi_freq_bifurc}.

\begin{figure}[H]
\vspace*{-5mm}
\centerline{\includegraphics[width=7.5cm,clip]{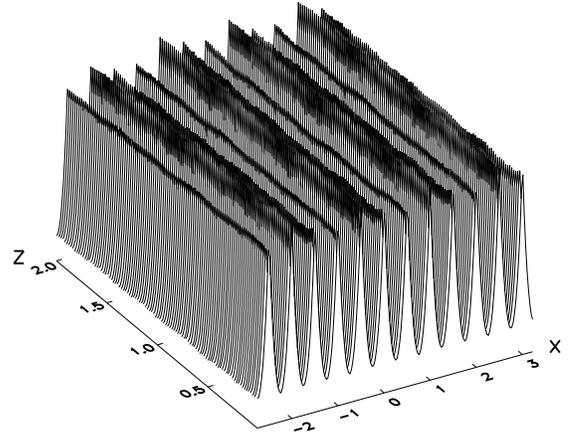}}
\vspace*{1mm}
\caption{ \label{fig:mi_bpm_unstag}
Development of modulational instability in a self-focusing medium for a perturbed unstaggered nonlinear BW mode with $I_0 \simeq 0.44$ ($\alpha=3$, $h=0.5$, and $\gamma=+1$.
}
\end{figure}

Modulational instability of the nonlinear Bloch waves in a periodic medium has been earlier studied for the Bose-Einstein condensates in optical lattices~\cite{mi_bec} in the mean-field approximation based on the Gross-Pitaevskii equation, which is mathematically equivalent to Eq.~(\ref{eq:nls}) with ${\cal F}(I; x) = \nu(x) + \gamma I$. It was demonstrated that the {\em unstaggered modes are always modulationally unstable in a self-focusing medium} ($\chi={\rm sign}\gamma=+1$), and they are {\em stable in a self-defocusing medium} ($\chi=-1$). It can be shown that the similar results are also valid for our model. Indeed, since the unstaggered waves are the fundamental modes of the self-induced periodic potential, oscillatory instabilities can not occur, and MI can only correspond to purely imaginary eigenvalues $\Gamma$. Such an instability should appear at the critical points defined by Eq.~(\ref{eq:mi_bif}) at $\Gamma=0$, which are found as $Q = 1, \eta+3$. Therefore, the range of the unstable modulation frequencies is
\[
 \begin{array}{l}
  {\displaystyle
     0 < q < \cos^{-1}( \eta+3 ), \quad -4 \le \eta < -2, 
  } \\*[9pt] {\displaystyle
     0 < q \le \pi, \quad \eta < -4, 
  } \end{array}
\]
According to Eq.~(\ref{eq:BW_I0}), for unstaggered modes we have $\chi \eta < -2 \chi$, and the (in)stability results follow immediately.
We note that at small intensities (when $\eta \simeq -2$) the modulational instability in a self-focusing medium corresponds to long-wave modulations, as illustrated in Fig.~\ref{fig:mi_bpm_unstag}.

The {\em staggered BW modes in a self-defocusing medium ($\chi=-1$) are also modulationally unstable}~\cite{mi_bec}. This happens because the staggered waves experience effectively ``normal'' diffraction~\cite{dm}. Such waves exist for $\eta>2$ and, similar to the case of unstaggered waves in a self-focusing medium, we identify the range of unstable frequencies corresponding to the purely imaginary eigenvalues (${\rm Re}\,\Gamma=0$), 
\[
 \begin{array}{l}
  {\displaystyle
     0 < q < \cos^{-1}( 3-\eta ), \quad 2 < \eta \le 4 , 
  } \\*[9pt] {\displaystyle
     0 < q \le \pi, \quad \eta > 4 .
  } \end{array}
\]

\begin{figure}[H]
\centerline{\includegraphics[width=7.5cm,clip]{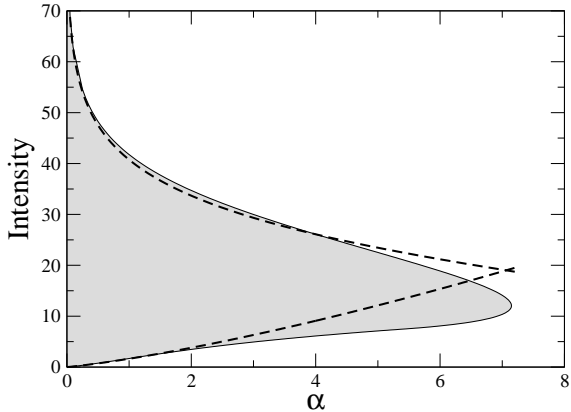}}
\vspace*{1mm}
\caption{ \label{fig:mi_int}
Modulationally unstable staggered BW modes (gray shading) in a self-focusing medium, shown as the intensity $I_0$ vs. the parameter $\alpha$ (at $h=0.5$, $\gamma=+1$). Dashed lines are the analytical approximations~(\ref{eq:mi_Icr}) and~(\ref{eq:mi_Icr_max}) for the low- and high-intensity instability thresholds, respectively.
}
\end{figure}

Finally, we analyze stability of {\em staggered BW modes in a self-focusing medium} ($\chi=+1$). Since such modes exist for $\eta<2$, the domain $1 < Q(\Gamma) < 3-\eta$ at ${\rm Re}\,\Gamma=0$ does not correspond to physically possible modulation frequencies. However, the oscillatory instabilities (i.e. those with complex $\Gamma$) can appear due to resonances between the modes that belong to different bands. Such instabilities appear in a certain region of the wave intensities in the case of shallow modulations, below a certain threshold value (when $\alpha h \le 3.57\ldots$), as shown in Fig.~\ref{fig:mi_int}.
We find the following asymptotic expression for the low-intensity instability threshold, 
\begin{equation} \label{eq:mi_Icr}
  \gamma I_0^{(min)} \simeq \alpha 
                   + 2 \sqrt{2 h} \alpha^{3/2} / \pi 
                   + O(\alpha^{2}) ,
\end{equation}
while the upper boundary is given by the relation
\begin{equation} \label{eq:mi_Icr_max}
  \alpha \simeq 4 \gamma I_0^{(max)} \exp( \gamma I_0^{(max)}h / 4 ) .
\end{equation}
These analytical estimates are shown with the dashed lines in Fig.~\ref{fig:mi_int}.

\begin{figure}[H]
\centerline{\includegraphics[width=7.5cm,clip]{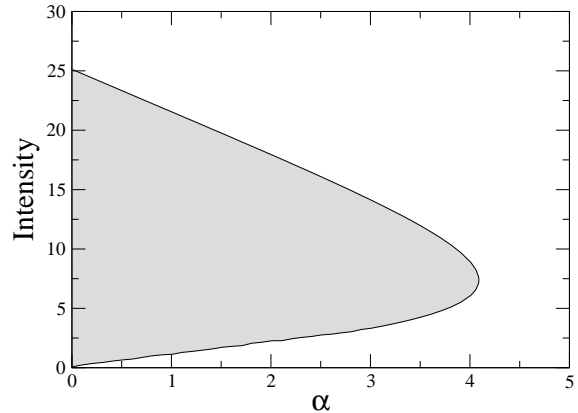}}
\vspace*{1mm}
\caption{ \label{fig:mi_int_2atom}
Region of modulational instability (gray shading) for the model~(\ref{eq:2atom}); notations are the same as in Fig.~\ref{fig:mi_int}. The result is only qualitatively similar to Fig.~\ref{fig:mi_int}, since the two-component discrete model~(\ref{eq:2atom}) is asymptotically correct for small intensities only.
}
\end{figure}

\begin{figure}[H]
\centerline{\includegraphics[width=7.5cm,clip]{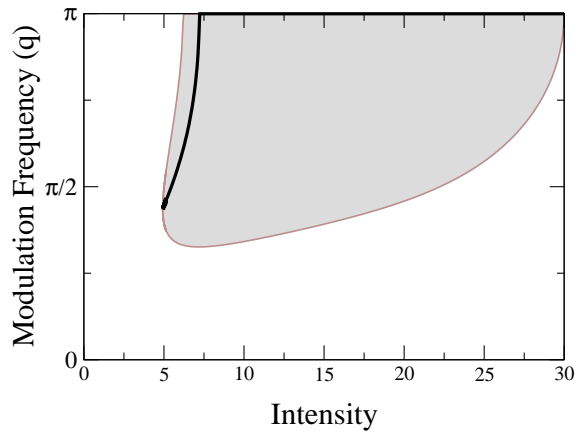}}
\vspace*{1mm}
\caption{ \label{fig:mi_freq}
Unstable modulation frequencies (gray shading) vs. intensity $I_0$, for  the staggered BW modes ($\alpha=3$, $h=0.5$, $\gamma=+1$). Solid line shows the parameters of the linear eigenmode with the largest instability growth rate (i.e. $\max |{\rm Im} \Gamma|$).
}
\end{figure}

It is interesting to compare our results with those obtained in the framework of the continuous coupled-mode theory (see Sec.~\ref{sect:coupled}), valid for the case of a narrow band gap (i.e. for small $\alpha$ and small $I_0$). Although the nonlinear coupling coefficients in Eq.~(\ref{eq:coupled}) are different compared to a couped-mode model for shallow gratings~\cite{mi_bragg}, the key stability result remains the same, and the oscillatory instability appears above a certain critical intensity proportional to the band-gap width. In our case, the band-gap width is $2 \alpha / h + O(\alpha^{3/2})$, and we observe a good agreement with the results of the coupled-mode theory.
However, the coupled-mode model~(\ref{eq:coupled}) can not predict the stability region at high intensities, since in this region the approximation is no longer valid.

In the limit of large $\alpha$, the BW dynamics can be studied with the help of the tight-binding approximation (see Sec.~\ref{sect:tight}). Then, the effective discrete NLS equation~(\ref{eq:dnls_tight}) predicts the stability of the staggered modes in a self-focusing medium~\cite{mi_dnls}. Numerical and analytical results confirm that our solutions are indeed stable in the corresponding parameter region.

The two-component discrete model introduced in Sec.~\ref{sect:superlattice} predicts the existence of oscillatory instabilities of the BW waves for $\beta$ satisfying the inequality
\[
    \frac{1}{2} (\beta-\beta_2) 
    - \frac{(\beta-\beta_2)^3}{32 \rho_1 \rho_2} - (\beta_1-\beta_2) > 0 ,
\]
where the coefficients $\rho_{1,2}$ are defined in Eq.~(\ref{eq:2atom-param}). The corresponding instability region in the parameter space intensity vs. lattice depth ($\alpha$) can be calculated using Eqs.~(\ref{eq:BW_I0}) and~(\ref{eq:2atom_dispers}), and the result is presented in Fig.~\ref{fig:mi_int_2atom}. Since the simplified model is (asymptotically) correct only for small intensities (i.e. when $\beta \simeq \beta_1$), there is no quantitative agreement between Figs.~\ref{fig:mi_int} and~\ref{fig:mi_int_2atom}. However, the model~(\ref{eq:2atom}) does predict the key pattern of oscillatory MI: (i)~instability appears only for a finite range of intensities when the grating depth ($\alpha$) is below a critical value, and (ii)~MI is completely suppressed for large $\alpha$. Thus, unlike the DNLS equation, the two-component discrete model~(\ref{eq:2atom}) predicts qualitatively all major features of MI in a periodic medium.

\begin{figure}[H]
\vspace*{-5mm}
\centerline{\includegraphics[width=7.5cm,clip]{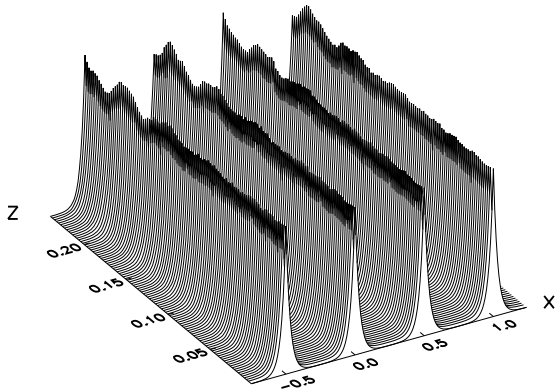}}
\vspace*{1mm}
\caption{ \label{fig:mi_bpm_stag}
Development of the instability-induced period-doubling modulations. Initial profile corresponds to a slightly perturbed staggered mode with $I_0 \simeq 29.87$. Parameters are the same as in Fig.~\ref{fig:mi_freq}.
}
\end{figure}

We find that in a self-focusing medium the staggered waves ($K=\pi$) are always stable with respect to low-frequency modulations, see Fig.~\ref{fig:mi_freq}. However, at larger intensities unstable frequencies appear close to the middle and the edge of the Brillouin zone, and they are shifted towards the edge, $q=\pi$, having there the largest instability growth rate. The corresponding modulational instability manifests itself through the development of the period-doubling modulations, as shown in Fig.~\ref{fig:mi_bpm_stag}.

\section{Bright Spatial Solitons} \label{sect:bright}
\subsection{Odd and Even Localized Modes} \label{sect:bright_slv}

Stationary localized modes in the form of discrete bright solitons can exist with the propagation constant inside the band gaps, when $|\eta|>2$. Additionally, such solutions can exist only if the nonlinearity and dispersion sign are different, i.e. when $\eta \chi < 2$. It follows from Eq.~(\ref{eq:dnls_param}), that $\beta>-(\pi/h)^2$ and $\xi > 0$ 
in the IR gap and the first BR gap (see also Fig.~\ref{fig:dispers}), so that the type of the nonlinear response is fixed by the medium characteristics, since $\chi = {\rm sign} (\gamma)$. Therefore, self-focusing nonlinearity can support bright solitons in the IR region (where $\eta<-2$), i.e. in the conventional {\em wave-guiding regime}. In the case of the self-defocusing response,  bright solitons can exist in the first BR gap, owing to the fact that the sign of the effective diffraction is inverted ($\eta>2$). In the latter case, the mode localization occurs in the so-called {\em anti-waveguiding regime}. 

Let us now consider the properties of two basic types of the localized modes: {\em odd}, centered at a nonlinear thin-film waveguide, and {\em even}, centered between the neighboring waveguides, so that $U_{|n|} = \chi^s U_{-|n|-s}$, where $s=0,1$, respectively.  For discrete lattices, such solutions have already been studied in the literature~(see, e.g., Ref.~\cite{Campbell}), and it has been  found that the mode profile is ``unstaggered'' (i.e. $U_n>0$) if $\eta<-2$. On the other hand, Eq.~(\ref{eq:dnls}) possesses a symmetry, 
\begin{equation} \label{eq:dnls_symm}
  U_n \rightarrow (-1)^n U_n,\quad
  \eta \rightarrow -\eta,\quad 
  \chi \rightarrow -\chi, 
\end{equation}
which means that the solutions become ``staggered'' at $\eta>2$. 
Because of this symmetry, it is sufficient to find localized solutions of Eq.~(\ref{eq:dnls}) for $\eta < -2$ and $\chi = +1$. The known approximate solutions give accurate results only in the case of highly localized modes ($|\eta|\gg 2$)~(see \cite{lederer,malomed}, and references therein) and in the continuous limit ($|\eta| \simeq 2$) (see, e.g.,~\cite{theory1b}). In order to describe the mode profile for arbitrary values of $\eta$, we introduce {\em a new approach based on the physical properties of localized solutions}. We recall that the nonlinear localized modes are similar to the impurity states, which explains the presence of a sharp central peak (or two peaks) in the mode profiles at large $\eta$. On the other hand, we have found that {the tails of a localized mode are always quite smooth}, both in the continuous limit and highly-discrete case. Based on these facts, we construct the approximate solutions by matching the mode tails with the central impurity node(s).

First, we have to find an approximation for the mode tails. Since the tail is smooth, its profile can be well approximated by the continuous equations. The simplest yet effective approach is to choose the model coefficients to match the discrete solutions at the beginning of the tail, which we define as the {\em zero concavity point}, and in the {\em linear limit} corresponding to the far-field asymptotics. Then, after simple calculations, we obtain an {\em approximate continuous equation for the mode tails},
\begin{equation} \label{eq:bright_tail}
  -\lambda U + \frac{\lambda}{\rho^2} \frac{d^2 U}{d n^2} + U^3 = 0 ,
\end{equation}
where $\lambda = -(\eta + 2) > 0$, and $\rho = {\rm cosh}^{-1}(1+\lambda/2)$. The solution of Eq.~(\ref{eq:bright_tail}) with the vanishing far-field asymptotics has the form
\begin{equation} \label{eq:bright_sech}
   U(n;\,n_s) = \sqrt{2 \lambda}\, {\rm sech}( \rho (n+n_s) ),
\end{equation}
where the free parameter $n_s$ defines the shift. Note that in the limit $\lambda \rightarrow 0$ we have $\rho \rightarrow \sqrt{\lambda}$, and the results of the conventional continuous approximation~\cite{theory} are recovered.

\begin{figure}[H]
\centerline{\includegraphics[width=7.5cm,clip]{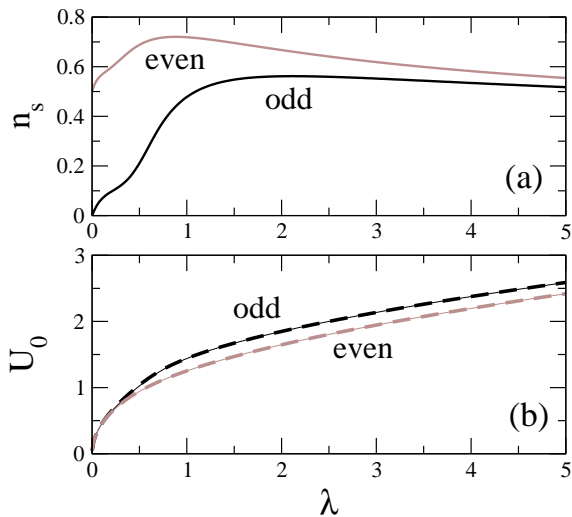}}
\vspace*{3mm}
\caption{ \label{fig:bright_xs}
Dependence of (a)~the shift parameter $n_s$ and (b)~peak normalized amplitude $U_0$ in Eqs.~(\ref{eq:tail_xs}),~(\ref{eq:bright_U0}) on the parameter $\lambda$ for odd and even bright localized modes. Dashed lines in (b)~--- numerically calculated values.
}
\end{figure}

\begin{figure}[H]
\centerline{\includegraphics[width=7.5cm,clip]{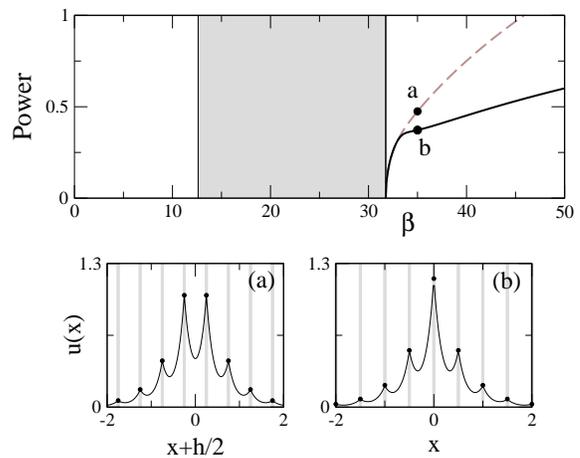}}
\vspace*{1mm}
\caption{ \label{fig:pwr_sf}
Top: power vs. propagation constant for odd (black) and even (dashed gray) localized modes in a self-focusing ($\gamma=+1$) regime: solid~--- stable, dashed~--- unstable.
Gray shading marks the transmission band.
Bottom:~profiles of the localized modes corresponding to the marked points (a) and (b) in the top plot; black dots~--- analytical approximation for the node amplitudes.
The lattice parameters are the same as in Fig.~\ref{fig:dispers}.
}
\end{figure}

\begin{figure}[H]
\centerline{\includegraphics[width=7.5cm,clip]{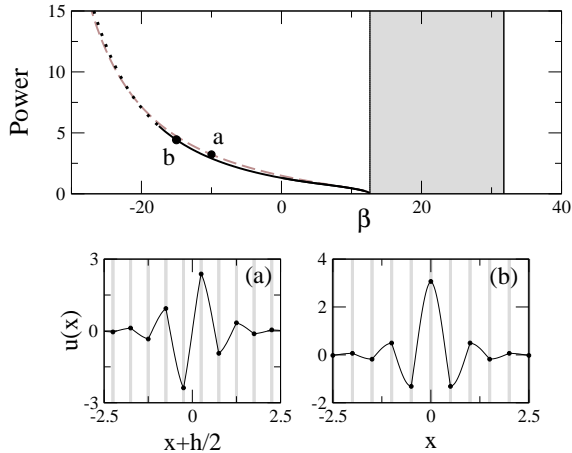}}
\vspace*{1mm}
\caption{ \label{fig:pwr_df}
Top: power vs. propagation constant in the self-defocusing ($\gamma=-1$) regime.
Dotted~--- oscillatory unstable modes; 
other notations are the same as in Fig.~\ref{fig:pwr_sf}.
}
\end{figure}

Second, we have to construct the full approximate solution as a combination of two tails. Such a state is supported by the central node in odd modes ($U_0$), and by two nodes in the case of even topology ($U_{-1}=\chi U_0$). Since the profiles are symmetric, we have to calculate the field structure only for $n \ge 0$, and the discrete tail profiles for $n \ge 1$ can be approximated as:
\begin{equation} \label{eq:tail_xs}
  U_n = U( n;\; n_s ) ,
\end{equation}
where the function $U( n;\; n_s )$ is given by Eq.~(\ref{eq:bright_sech}). In order to determine the unknown parameters, the peak amplitude $U_0$ and shift $n_s$, we should solve the original discrete equations~(\ref{eq:dnls}) at the soliton peak ($n=0$) and at the neighboring site ($n=1$):
\begin{eqnarray} \label{eq:bright_U0}
 \begin{array}{l}
  {\displaystyle 
     - (2 - s + \lambda ) U_0 + (2-s) U_1 + U_0^3 = 0 ,
  } \\*[9pt] {\displaystyle 
      - (2 + \lambda) U_1 + U_0 + U_2 + U_1^3 = 0,
  } \end{array}
\end{eqnarray}
where we took into account the symmetry properties of odd and even modes. We find that for all $\lambda>0$ there exists a solution that belongs to a smooth branch, starting with $n_s = s/2$ in the continuous limit (at $\lambda \rightarrow 0$), see Fig.~\ref{fig:bright_xs}(a). We compare the analytical approximation with the exact numerical solution of the original model~(\ref{eq:dnls}) for the peak amplitude $U_0$, and find that an error  does not exceed 1.5\% for odd and 0.8\% for even modes. As a matter of fact, the corresponding curves are indistinguishable in Fig.~\ref{fig:bright_xs}(b).
The actual mode profiles are also adequately represented, see examples in Figs.~\ref{fig:pwr_sf}(a,b) and~\ref{fig:pwr_df}(a,b). Therefore, the suggested analytical procedure allows us to obtain extremely accurate approximate analytical solutions in the whole parameter range, including the extreme cases of the continuous ($\lambda \rightarrow 0$) and anti-continuous ($\lambda \rightarrow +\infty$) limits.

\begin{figure}
\vspace*{5mm}
\centerline{\includegraphics[width=7.5cm,clip]{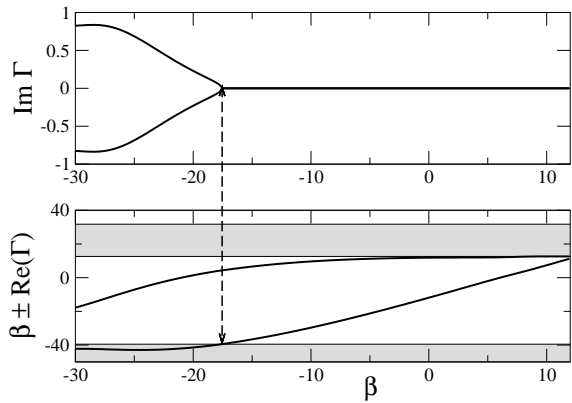}}
\vspace*{5mm}
\caption{ \label{fig:imode_df_odd}
Example of a resonance between an eigenmode of the linear eigenvalue problem and a bang-gap edge of the continuous spectrum that leads to an oscillatory instability of the odd localized mode (parameters correspond to Fig.~\ref{fig:pwr_df}).}
\end{figure}

Our linear stability analysis reveals that even modes are {\em always unstable} with respect to a translational shift along the $x$ axis. On the other hand, odd modes are always stable in the self-focusing regime (see Fig.~\ref{fig:pwr_sf}, top), but can exhibit {\em oscillatory instabilities} in the self-defocusing case when the power exceeds a certain critical value (see Fig.~\ref{fig:pwr_df}, top). At this point, an eigenmode of the linearized problem resonates with the band-gap edge, the value $(\beta-{\rm Re}\;\Gamma)$ moves inside the band, and non-zero imaginary part of the eigenvalue appears, as illustrated by an example in Fig.~\ref{fig:imode_df_odd}. Such an instability scenario is similar to one earlier identified for gap solitons~\cite{gap_inst}, and also for the modes localized at a single nonlinear layer in a linear periodic structure~\cite{our_PRL}. The latter example demonstrates a deep similarity between the periodic systems with localized and distributed nonlinearities.

\begin{figure}[H]
\vspace*{-15mm}
\centerline{\includegraphics[width=7.5cm,clip]{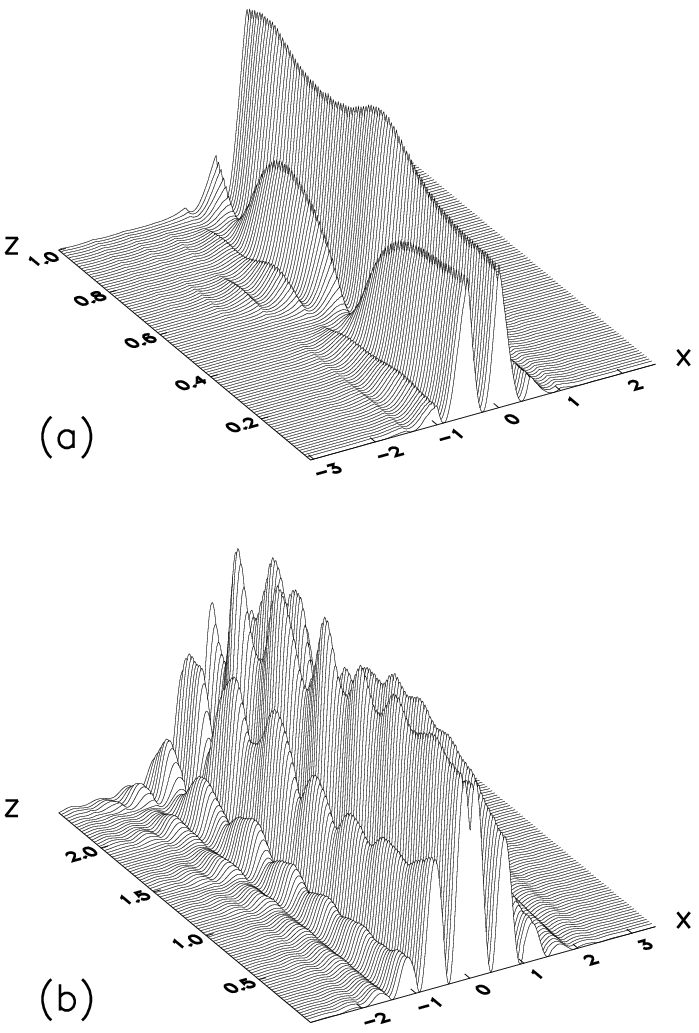}}
\vspace*{1mm}
\caption{ \label{fig:bright_df_bpm}
Instability development for the Bragg-type localized modes in a self-defocusing medium:
(a)~symmetry-breaking instability of an even mode ($P \simeq 3.23$);
(b)~oscillatory instability of an odd mode ($P \simeq 23.4$). 
Parameters correspond to Fig.~\ref{fig:pwr_df}.
}
\end{figure}

Development of both types of instability is demonstrated in Figs.~\ref{fig:bright_df_bpm}(a,b). Transformation of an even mode into an odd counterpart due to a symmetry-breaking instability is shown in Fig.~\ref{fig:bright_df_bpm}(a). The effect of oscillatory instability on an odd mode is quite different: strong radiation is emitted due to a resonant coupling with linear waves outside the band gap, see Fig.~\ref{fig:bright_df_bpm}(b).

\subsection{Soliton Bound States~--- ``Twisted'' Modes} 
            \label{sect:bright_bs}

Due to a periodic modulation of the medium refractive index, solitons can form bound states~\cite{bound}. In particular, the so-called {\em ``twisted'' localized mode}~\cite{twisted} is a combination of two out-of-phase bright solitons~\cite{bound}. Such solutions do not have their continuous counterparts, and they can only exist when the discreteness effects are strong, i.e. for $|\eta| > \eta_{\rm cr}$. Properties of the twisted modes depend on the separation between the modes forming a bound state. We consider the cases of two lowest-order solutions of (i)~``even'' type with zero nodes ($m=0$) in-between the peaks, and (ii)~``odd'' type with one node ($m=1$) at the middle, with $U_0 \equiv 0$. The corresponding symmetry properties are $U_{|n|+m} = - \chi^{m+1} U_{-|n|-1}$. Additionally, Eq.~(\ref{eq:dnls_symm}) also holds, so that we only have to construct solutions for $\chi=+1$. Then, the and the soliton tails at $n > m$ are approximated as:
\begin{equation} \label{eq:tail_tw_xs}
  U_n = U( n - m; \; n_s ) ,
\end{equation}
where $U(n;\; n_s)$ is given by Eq.~(\ref{eq:bright_sech}). The matching conditions are:
\begin{eqnarray} \label{eq:bright_tw_U0}
 \begin{array}{l}
  {\displaystyle 
     - (3 - m + \lambda ) U_m + U_{m+1} + U_m^3 = 0 ,
  } \\*[9pt] {\displaystyle 
      - (2 + \lambda) U_{m+1} + U_m + U_{m+2} + U_{m+1}^3 = 0,
  } \end{array}
\end{eqnarray}
where, as before, $\lambda = -(\eta + 2) > 0$.

\begin{figure}[H]
\centerline{\includegraphics[width=7.5cm,clip]{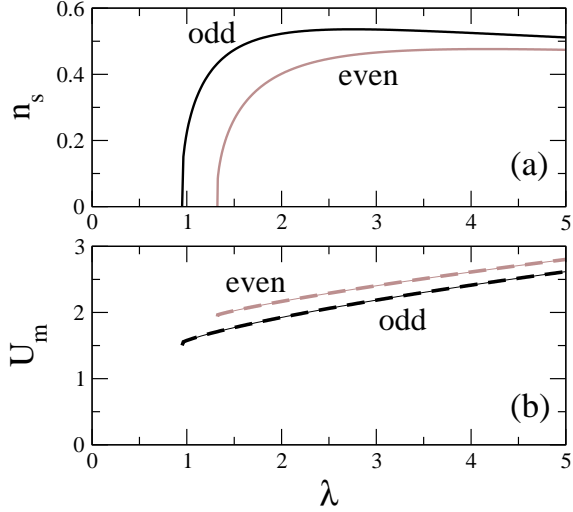}}
\vspace*{3mm}
\caption{ \label{fig:bright_tw_xs}
Dependence of (a)~the shift parameter $n_s$ and (b)~the peak normalized amplitude $U_m$ in Eqs.~(\ref{eq:tail_tw_xs}) and~(\ref{eq:bright_tw_U0}) on the parameter $\lambda$, for odd and even twisted localized modes. Dashed lines in (b)~--- numerically calculated values.
}
\end{figure}

\begin{figure}[H]
\centerline{\includegraphics[width=7.5cm,clip]{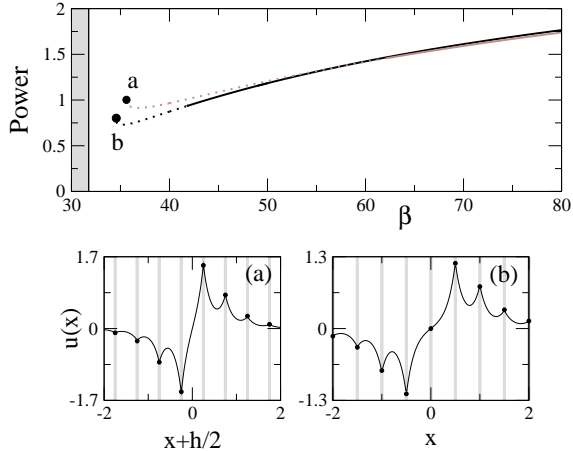}}
\vspace*{1mm}
\caption{ \label{fig:pwr_sf_bright_tw}
Top: power vs. propagation constant for odd (black) and even (gray) twisted localized waves in a self-focusing ($\gamma=+1$) regime.
Notations are the same as in Figs.~\ref{fig:pwr_sf} and~\ref{fig:pwr_df}.
}
\end{figure}

\begin{figure}[H]
\centerline{\includegraphics[width=7.5cm,clip]{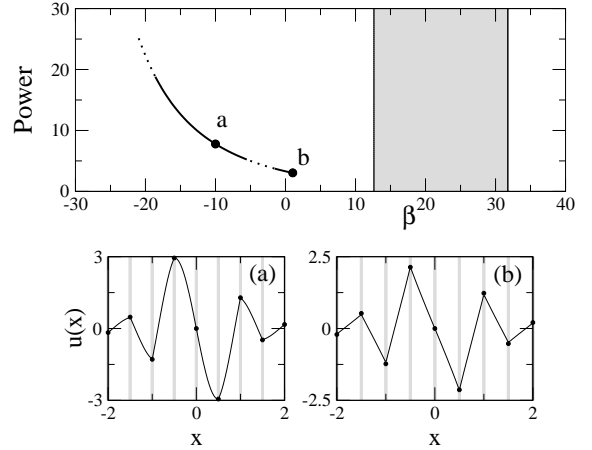}}
\vspace*{1mm}
\caption{ \label{fig:pwr_df_bright_tw}
Top: power vs. propagation constant for the odd twisted localized waves in a self-defocusing ($\gamma=-1$) regime.
Notations are the same as in Figs.~\ref{fig:pwr_sf} and~\ref{fig:pwr_df}.
}
\end{figure}

We determine solution of Eqs.~(\ref{eq:bright_tw_U0}) starting with the highly localized modes earlier described anti-continuous limit ($\lambda \gg 1$)~\cite{twisted}, and then gradually decrease the parameter $\lambda$. We find that solution exists for $\lambda > \lambda_{\rm cr} >0$, and disappears when $n_s(\lambda_{\rm cr}) \equiv 0$. Substituting this condition into 
Eq.~(\ref{eq:bright_tw_U0}), we determine the approximate critical parameter values, $\eta_{\rm cr}(m=0) \simeq 3.32$ and $\eta_{\rm cr}(m=1) \simeq 2.95$. The {\em analytically determined existence regions agree very well with numerical results} (within 1\%). To the best of our knowledge, none of the previously developed analytical approximations could predict the regions of existence for highly discrete twisted modes (with small $m$). Moreover, the approximate solution describes very accurately the profiles of the twisted modes, see examples in Figs.~\ref{fig:pwr_sf_bright_tw}(a,b) and Figs.~\ref{fig:pwr_df_bright_tw}(a,b). In particular, a relative error for the peak amplitude $U_m$ is less than 0.5\%, so that in Fig.~\ref{fig:bright_tw_xs}(b) the analytical and numerical dependencies practically coinside.

In the self-focusing regime, stability properties of the twisted modes in the IR gap can be similar to those earlier identified in the framework of a DNLS model~\cite{twisted}. In the example shown in  Fig.~\ref{fig:pwr_sf_bright_tw}, the modes are stable at larger values of the propagation constant, and they become oscillatory unstable closer to the boundary of the existence region. Quite importantly, the stability region is much wider in the case of odd twisted modes due to a larger separation between the individual solitons of the bound state.

\begin{figure}[H]
\vspace*{-15mm}
\centerline{\includegraphics[width=7.5cm,clip]{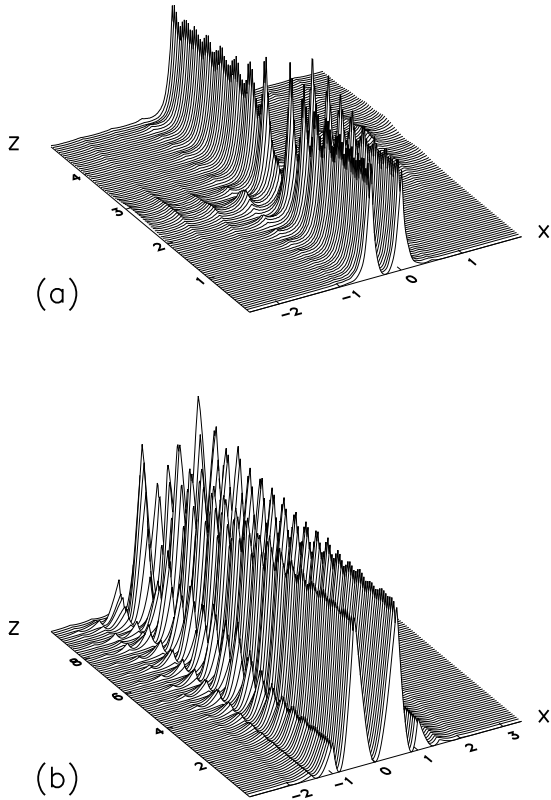}}
\vspace*{1mm}
\caption{ \label{fig:bright_tw_bpm}
Instability scenarios for twisted localized modes:
(a)~an even mode in a self-focusing medium ($P \simeq 1.22$):
(b)~an odd mode in a self-defocusing regime ($P \simeq 4.45$).
Parameters correspond to Figs.~\ref{fig:pwr_sf_bright_tw} and~\ref{fig:pwr_df_bright_tw}, respectively.
}
\end{figure}

The characteristics of the twisted modes in the BR gap can differ substantially from the previous case. First, the value of $\eta$ is limited from above ($2 < \eta <\eta_{\rm max}$) and, therefore, some families of the twisted modes with $m < m_{\rm cr}$ may not exist. For example, for the medium parameters corresponding to Fig.~\ref{fig:dispers}(a), we have 
$\eta_{\rm cr}(m=1) < (\eta_{\rm max} \simeq 3.18 ) < \eta_{\rm cr}(m=0)$, 
so that $m_{\rm cr} = 1$. Under these conditions, even modes with $m=0$ cannot exist in the BR regime. On the other hand, the odd modes with $m=1$ can exist, and they are stable in a wide parameter region, see Fig.~\ref{fig:pwr_df_bright_tw}. Development of oscillatory instabilities for IR and BR twisted modes is illustrated in Figs.~\ref{fig:bright_tw_bpm}(a) and~\ref{fig:bright_tw_bpm}(b), respectively. In both the cases, we observe an exponential increase of periodic amplitude modulations and the emission of radiation waves.

\section{Dark Spatial Solitons} \label{sect:dark}

Similar to the continuous NLS equation with self-defocusing nonlinearity~\cite{dark_review} or the DNLS equation~\cite{dark_inst}, our model can support {\em dark solitons}~--- localized modes on the Bloch-wave background. However, dark stationary localized modes in a periodic medium can exist for both signs of nonlinearity. To be specific, let us consider the case of a background corresponding to the Bloch-wave solutions with $K=0,\pi$ introduced in Sec.~\ref{sect:array_mi}. Then, dark-mode solutions
can appear at the band-gap edge where $\eta = 2 \chi = 2 {\rm sign} \gamma$, since in such a case nonlinear and dispersion terms have the same signs~\cite{dark_review}.

Similar to the case of bright solitons discussed above, two basic types of dark spatial solitons can be identified, namely,  {\em odd localized modes} centered at a nonlinear thin-film waveguide, and {\em even localized modes} centered between the neighboring thin-film waveguides. All such modes satisfy the symmetry condition, $U_{|n|+s} = - (-\chi)^{s+1} U_{-|n|-1}$, where $s=0,1$ for even and odd modes, respectively. The BW background is unstaggered if $\chi = -1$, and it is staggered for $\chi = +1$; the corresponding solutions  can be constructed with the help of a symmetry transformation, $U_n \rightarrow (-1)^n U_n$. However, the stability properties of these two types of localized states can be quite different. Indeed, it has been demonstrated in Sec.~\ref{sect:array_mi} that in a self-focusing medium ($\chi=+1$) the staggered background can become unstable. On the contrary, the unstaggered background is always stable if $\chi=-1$.

In order to find the approximate analytical solutions, we consider the case $\chi=-1$, with no lack of generality [since solutions with $\chi=+1$ can be obtained by applying the symmetry transformation~(\ref{eq:dnls_symm})]. In this case, the far-field asymptotics for solutions of Eq.~(\ref{eq:dnls}) close to the background level can be found as $(U_\infty-U_n) \simeq e^{\rho n}$, where $U_\infty = \sqrt{\lambda}$ is the background amplitude, $\rho = {\rm cosh}^{-1}(1+\lambda)$ is the localization parameter, and $\lambda = \eta +2 > 0$. Then, we obtain an approximate continuous equation for the nonlinear mode tails by matching the asymptotic solution at large~$n$,
\begin{equation} \label{eq:dark_tail}
   \lambda U + \frac{2 \lambda}{\rho^2} \frac{d^2 U}{d n^2} - U^3 = 0 .
\end{equation}
The corresponding dark-soliton solution has the form
\begin{equation} \label{eq:dark_tanh}
   U(n;\; n_s) = \sqrt{\lambda}\, {\rm tanh}( \rho (n+n_s) / 2 ).
\end{equation}
Note that in the limit $\lambda \rightarrow 0$ we have $\rho \rightarrow \sqrt{2 \lambda}$, and the results of the conventional continuous approximation are recovered.

\begin{figure}[H]
\centerline{\includegraphics[width=7.5cm,clip]{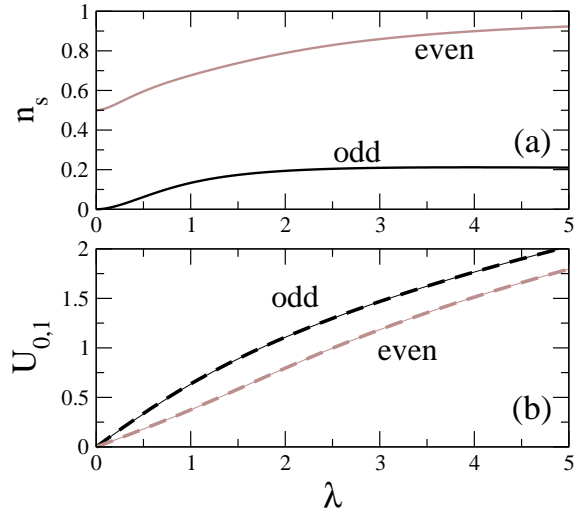}}
\vspace*{3mm}
\caption{ \label{fig:dark_xs}
Dependence of (a)~the shift parameter $n_s$ and (b)~the normalized amplitudes $U_s$ in Eqs.~(\ref{eq:tail_xs}),(\ref{eq:dark_tanh}), and~(\ref{eq:dark_U0}) on the parameter $\lambda$ for odd ($s=1$) and even ($s=0$) twisted localized modes. Dashed lines in (b)~--- numerically calculated values.
}
\end{figure}

Similar to the discrete bright solitons, a localized solution can be constructed by matching the soliton tails defined by Eq.~(\ref{eq:tail_xs}). The corresponding matching conditions,
\begin{eqnarray} \label{eq:dark_U0}
 \begin{array}{l}
  {\displaystyle 
     (\lambda - 2 - s) U_0 + (1-s) U_1 - U_0^3 = 0 ,
  } \\*[9pt] {\displaystyle 
      (\lambda - 2) U_1 + U_0 + U_2 - U_1^3 = 0 , 
  } \end{array}
\end{eqnarray}
are used to determine the shift parameter $n_s$ and amplitude $U_0$. We have $U_0 \equiv 0$ for odd modes, due to their symmetry properties.
Dependencies of the shift parameter $n_s$ on $\lambda$ are presented in Fig.~\ref{fig:dark_xs}(a). The soliton amplitudes at the central sites are shown in Fig.~\ref{fig:dark_xs}(b), where we observe again an excellent agreement between the approximate analytical and numerical solutions (the corresponding errors do not exceed 2\% for odd and 1\% for even modes).

\begin{figure}[H]
\centerline{\includegraphics[width=7.5cm,clip]{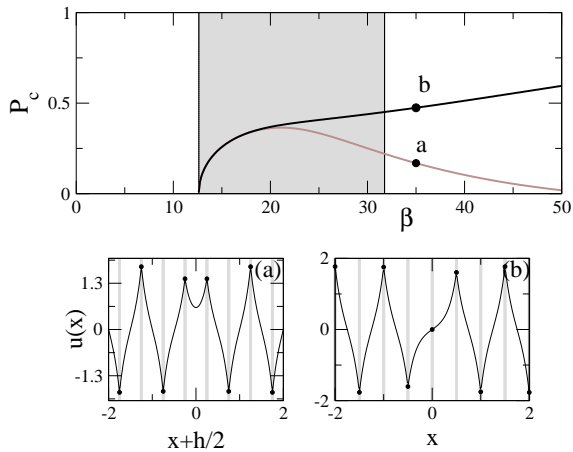}}
\vspace*{1mm}
\caption{ \label{fig:pwr_sf_dark}
Top: complementary power vs. propagation constant for odd (black) and even (gray) dark localized solitons in a self-focusing ($\gamma=+1$) regime.
Notations are the same as in Fig.~\ref{fig:pwr_sf}, but (in)stability regions are not indicated.
}
\end{figure}

\begin{figure}[H]
\centerline{\includegraphics[width=7.5cm,clip]{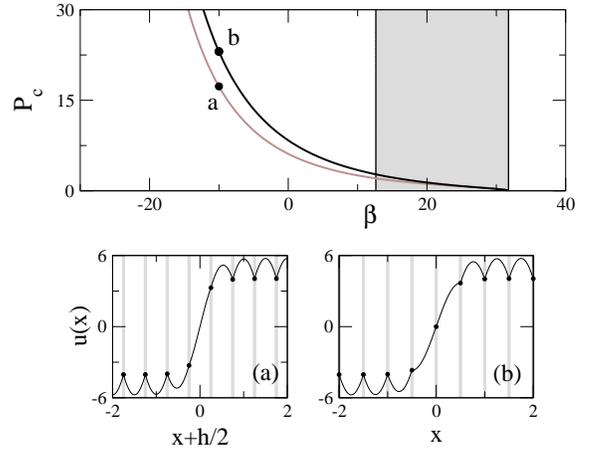}}
\vspace*{1mm}
\caption{ \label{fig:pwr_df_dark}
Top: complementary power vs. propagation constant in the self-defocusing ($\gamma=-1$) regime. Notations are the same as in Fig.~\ref{fig:pwr_sf_dark}.
}
\end{figure}

Two types of dark spatial solitons in our model are shown for both staggered and unstaggered BW backgrounds in Figs.~\ref{fig:pwr_sf_dark} and~\ref{fig:pwr_df_dark}, respectively. We characterize the family of dark solitons by the {\em complimentary power} defined as  
\[
  P_c = \lim_{n\rightarrow+\infty} \int_{-nh}^{+nh} 
         \left( |u(x+2 n h)|^2 - |u(x)|^2 \right) \; dx ,
\]
where $n$ is integer. 
The localized solutions shown in Fig.~\ref{fig:pwr_df_dark} are similar to those found earlier in Ref.~\cite{dark_NLS} in the context of the superflow dynamics on a periodic potential. 

\begin{figure}[H]
\centerline{\includegraphics[width=7.5cm,clip]{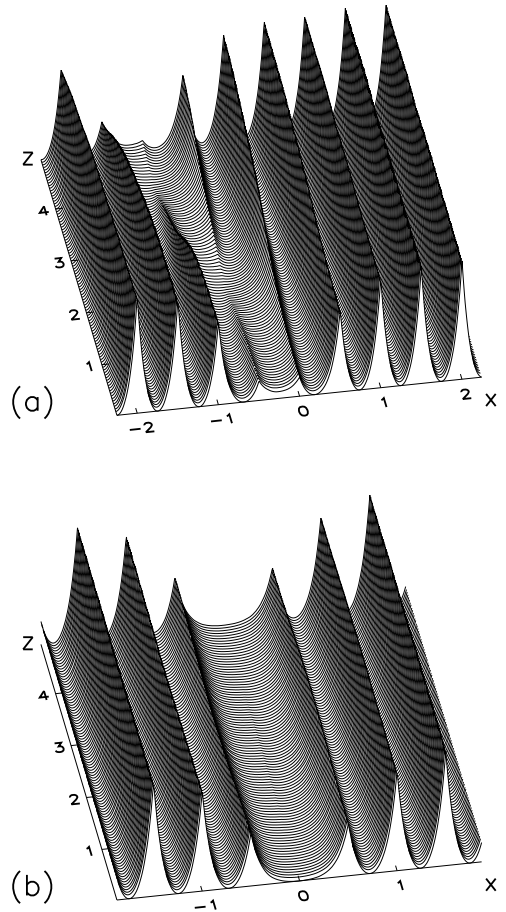}}
\vspace*{1mm}
\caption{ \label{fig:dark_sf_bpm}
Propagation dynamics of (a)~even and (b)~odd 
dark localized modes in a self-focusing medium. Initial profiles correspond to slightly perturbed stationary solutions at $\beta=20$, other 
parameters correspond to Fig.~\ref{fig:pwr_sf_dark}. }
\end{figure}

Numerical study of the propagation dynamics demonstrates that, similar to the case of bright solitons, even dark-soliton modes are unstable with respect to asymmetric perturbations, see an example in Fig.~\ref{fig:dark_sf_bpm}(a). On the other hand, odd modes can propagate in a stable (or weakly unstable) manner, as illustrated in Fig.~\ref{fig:dark_sf_bpm}(b). We note however that dark solitons can exhibit oscillatory instabilities close to the continuum limit~\cite{dark_inst} (at small intensities), but a detailed analysis of the dark-mode stability is beyond the scope of the present paper.

\section{Conclusion}

In the framework of a simplified model of a nonlinear layered medium that describes the so-called Dirac-comb nonlinear waveguide array, we have analyzed spatial optical solitons in the form of bright, dark, and ``twisted'' localized modes.  In general, such solitons are of two types, i.e. they are either (i)~nonlinear guided waves localized due to the total internal reflection or (ii)~the Bragg-type localized modes existing in the forbidden transmission gaps, gap solitons.  We have analyzed the existence and stability of the nonlinear localized modes of both types and described also modulational instability of extended modes induced by a periodic change of the medium refractive index.  Additionally, we have discussed both similarities and differences with the models described by the DNLS equation, derived in the frequently used tight-binding approximation, and with the results of the coupled-mode theory, which are valid for a shallow modulation and a narrow gap in the transmission spectrum.  We believe our analysis and results may be useful for other fields, such as the nonlinear dynamics of the Bose-Einstein condensates in optical lattices (see, e.g., Ref.~\cite{smerzi}).

\section*{Acknowledgments}

We are indebted to O.~Bang and C.~M.~Soukoulis for useful collaboration at the initial stage of this project, and to Y.~Silberberg and G.~I.~Stegeman for encouraging discussions and interest to this project. The work was supported by the Performance and Planning Fund of the Institute of Advanced Studies at the Australian National University, and by the Australian Photonics Cooperative Research Center.

\end{multicols}
\end{document}